# Upscaling of Solute Transport in Heterogeneous Media with Non-uniform Flow and Dispersion Fields

Zhijie Xu[1,a] and Paul Meakin[2,3,4]

1. Energy Resource Recovery & Management, Idaho National Laboratory, Idaho Falls, Idaho 83415 USA. Now at

Computational Mathematics Group, Fundamental and Computational Sciences Directorate, Pacific Northwest National Laboratory, Richland, WA 99352, USA

2. Carbon Resource Management Department, Idaho National Laboratory, Idaho Falls, Idaho 83415, USA

3. Center for the Physics of Geological Processes, University of Oslo, Norway

4. Institute for Energy Technology, Kjeller, Norway

**Abstract**

An analytical and computational model for non-reactive solute transport in periodic heterogeneous media with arbitrary non-uniform flow and dispersion fields within the unit cell of length $\varepsilon$ is described. The model lumps the effect of non-uniform flow and dispersion into an effective advection velocity $V_e$ and an effective dispersion coefficient $D_e$. It is shown that both $V_e$ and $D_e$ are scale-dependent (dependent on the length scale of the microscopic heterogeneity, $\varepsilon$), dependent on the Péclet number $P_e$, and on a dimensionless parameter $\alpha$ that represents the effects of microscopic heterogeneity. The parameter $\alpha$, confined to the range of [-0.5, 0.5] for the numerical example presented, depends on the flow direction and non-uniform flow and dispersion fields. Effective advection velocity $V_e$ and dispersion coefficient $D_e$ can be derived for any given flow and dispersion fields, and $\varepsilon$. Homogenized solutions describing the macroscopic variations can be obtained from the effective model. Solutions with sub-unit-cell accuracy can be constructed by homogenized solutions and its spatial derivatives. A numerical implementation of the model compared with direct numerical solutions using a fine grid, demonstrated that the new method was in good agreement with direct solutions, but with significant computational savings.

---

[a] Electronic mail: zhijie.xu@pnnl.gov Tel: 509-372-4885

## I. Introduction

Solute transport in quasi one-dimensional systems or in a propagating quasi-planar front is often described by the one-dimensional advection-dispersion equation,

$$\frac{\partial c}{\partial t} + \frac{\partial (Vc)}{\partial x} = \frac{\partial}{\partial x}\left( D \frac{\partial c}{\partial x} \right),$$ (1)

where $c(x,t)$ is the solute concentration at position $x$ and time $t$, $V$ is the advection velocity and $D$ is the dispersion coefficient. The dispersion equation describes the average concentration in the dispersion front, but significant concentration fluctuations, not described by the dispersion equation, may exist in the dispersion front, particularly at high Péclet numbers. For example, in high Péclet numbers dispersion experiments by Måløy *et al.* [1] found that Eq. (1) describes the average concentration field in a porous medium that is inhomogeneous on short length scales and homogeneous on long length scales, even though isoconcentration contours in the dispersion front were self-affine. The advection-dispersion equation (1) is used extensively in subsurface hydrology, chemical engineering, environmental science, and petroleum engineering to describe the spatial and temporal variation of non-reactive solute concentrations. In the simplest model for solute transport in homogeneous media with uniform flow and dispersion ($V$ and $D$ are constants), the linear one-, two-, and three-dimensional advection-dispersion equations can be solved with given initial and boundary conditions. For some simple cases, analytical solutions have been compiled in various books [2-4]. However, solute transport in non-uniform flow and dispersion fields, where both $V$ and $D$ are position and/or time dependent, is more relevant to most practical applications. For example, simulation of the migration of contaminants with ground water flow in variable aperture fracture and/or porous media often requires position and/or time dependent flow and dispersion coefficients.

A number of analytical solutions have been developed for time-dependent transport coefficients [5-7], and even more solutions for position-dependent transport coefficients can be found in the literature [8-11]. Most of these solutions assume a simple form (such as linear or exponential) for the time or position dependence of the transport coefficients in an infinite or semi-infinite heterogeneous media. The purpose of this paper is to introduce a novel homogenization method to investigate the effective properties of heterogeneous media with non-uniform flow and dispersion field based on our previous studies on the homogenization method development [12-15]. In this paper, the heterogeneous porous media is represented by a spatially periodic array with a unit cell of length $\varepsilon$, as shown in Fig. 1. The unit cell is a representative elementary control volume. The length scale $\varepsilon$, the scale of the microscopic heterogeneity, is assumed to be much smaller than the size $L$ of the computational domain ($\varepsilon \Box \ L$). Solute is transported along a model flow path with variable advection velocity $V(y)$ and dispersion coefficient $D(y)$, where a local coordinate $y$ defined as

$$y = \frac{x - x_0}{\varepsilon} \tag{2}$$

can be introduced inside the unit cell, and $x_0$ is the origin of the unit cell. The flow in the unit cell is assumed to have an arbitrary $y$-dependent advective velocity $V(y)$ and a $y$-dependent dispersion coefficient $D(y)$. We show that non-reactive solute transport in periodic permeable media can be described in terms of an effective advection velocity $V_e$ and an effective dispersion coefficient $D_e$, both of which are functions of the length scale $\varepsilon$, the velocity field $V(y)$ and the dispersion field $D(y)$ within the unit cell. The results are relevant to the upscaling of non-reactive solute transport in variable aperture fractures and porous media.

## II. Formulation of the Model

We start from the standard solute transport equation with uniform transport coefficients given in Eq. (1). The general solution has the Fourier form

$$c(x,t) = \frac{1}{2\pi}\int_{-\infty}^{\infty} F(k) e^{i(kx-\omega t)} dk \,,$$
(3)

where $k$ is the wavenumber, $\omega(k)$ is the $k$-dependent frequency, and $F(k)$ is the Fourier transform of the concentration field $c$. By substituting Eq. (3) into (1), the dispersion relationship

$$\omega = kV - ik^2 D$$
(4)

can be obtained. The next step is to derive a similar relationship for the $k$-dependent frequency $\omega$ for a periodic heterogeneous media with position-dependent transport coefficients using a Floquet-Bloch approach [16, 17]. Within the unit cell, the advection-dispersion equation has the form

$$\frac{\partial c}{\partial t} + \frac{\partial(V(y)c)}{\partial x} = \frac{\partial(D(y)\partial c/\partial x)}{\partial x} \,.$$
(5)

The concentration $c(x,t)$ can be expressed in a general Fourier form

$$c(x,t) = \int_{-\infty}^{\infty} c_k(x,t) dk = \frac{1}{2\pi}\int_{-\infty}^{\infty} F(k) A(k,y) e^{i(kx-\omega t)} dk \,,$$
(6)

where $c_k(x,t)$ is the concentration corresponding to the wavenumber $k$,

$$c_k(x,t) = \frac{1}{2\pi} F(k) A(k,y) e^{i(kx-\omega t)}. \tag{7}$$

$A(k,y)$ is the $k$- and $y$- dependent concentration amplitude within the unit cell. In this and subsequent equations, $A(y)$ is used instead of $A(k,y)$ to simplify the notation. Equation (6) can be reduced to Eq. (3), where $A(y)=1$ for constant velocity $V$ and dispersion fields $D$. $A(y)$ is used to describe the influence of microscopic heterogeneity [16, 17] and it should be a periodic function of ( $A(y) = A(y+1)$ ) for periodic heterogeneous fields. Based on Eq. (6), the first order derivatives with respect to $x$ and $t$ are given by

$$\frac{\partial c}{\partial x} = \frac{1}{2\pi} \int_{-\infty}^{\infty} F(k) \left( \frac{1}{\varepsilon} \frac{\partial A}{\partial y} + ikA \right) e^{i(kx-\omega t)} dk , \tag{8}$$

and

$$\frac{\partial c}{\partial t} = \frac{1}{2\pi} \int_{-\infty}^{\infty} F(k)(-i\omega) A e^{i(kx-\omega t)} dk . \tag{9}$$

Substitution of these derivatives into the advection-dispersion Eq. (5) leads to the relationship

$$\frac{\partial q}{\partial y} + ik\varepsilon q + i\varepsilon^2 \omega A = 0 , \tag{10}$$

for any wavenumber $k$, where $q(y)$ is the $k$- and $y$-dependent amplitude of the concentration flux defined as,

$$q = D\left( \frac{\partial A}{\partial y} + ik\varepsilon A \right) - \varepsilon V A . \tag{11}$$

Here both $D(y)$ and $V(y)$ are position-dependent fields within the unit cell. It is clear that $A$, $\omega$, and $q$ are all $k$-dependent quantities and they can be expanded as Taylor series with respected to the scale parameter $\varepsilon$ (with respect to wavenumber $k$ for the expansion of dispersion relation $\omega(k)$):

$$A = A_0 + \varepsilon A_1 + \varepsilon^2 A_2 + ..., \tag{12}$$

$$\omega = c_1 k + \varepsilon c_2 k^2 + ... , \tag{13}$$

and

$$q = q_0 + \varepsilon q_1 + \varepsilon^2 q_2 + \varepsilon^3 q_3 + \varepsilon^4 q_4 + ..., \tag{14}$$

where the expansion coefficients $c_n$ ($n$=1,2) in Eq. (13) , which will be needed later, are not position dependent (y-dependent) and can be related to the effective transport coefficients through a direct comparison between Eqs. (13) and (4), namely $c_1 = V_e$ and $\varepsilon c_2 = -i D_e$ , where $V_e$ and $D_e$ are effective advection velocity and dispersion coefficient. The relationships

$$q_0 = D \frac{\partial A_0}{\partial y} \ \text{ and } \ q_n = D \left( \frac{\partial A_n}{\partial y} + ik A_{n-1} \right) - V A_{n-1} \ \text{ for } \ n > 0 \tag{15}$$

between $q_n$ and $A_n$ can be obtained by substituting expansions (14) and (12) into Eq. (11). Substituting expansions (12)-(14) into Eq. (10) yields the following equations for $q_n$ for each order of $\varepsilon$ :

$$\varepsilon^0: \quad B_0 = \frac{\partial q_0}{\partial y} = 0 ; \tag{16}$$

$$\varepsilon^1: \quad B_1 = \frac{\partial q_1}{\partial y} + ik q_0 = 0 ; \tag{17}$$

$$\varepsilon^2: \quad B_2 = \frac{\partial q_2}{\partial y} + ik q_1 + i A_0 c_1 k = 0 ; \tag{18}$$

and

$$\varepsilon^3: \quad B_3 = \frac{\partial q_3}{\partial y} + ikq_2 + i\left(A_1c_1k + A_0c_2k^2\right) = 0 \,. \tag{19}$$

The boundary conditions at the edges of the unit cell must be considered. The concentration continuity condition at the boundary (due to the periodicity) requires that

$$A(y = 0) = A(y = 1) \,. \tag{20}$$

By using the expansion (12), this relationship leads to

$$A_n(y = 0) = A_n(y = 1) \text{ for } n \geq 0 \,. \tag{21}$$

Similarly, the concentration flux continuity condition requires the flux amplitude $q_n$ to satisfy

$$q_n(y = 0) = q_n(y = 1) \text{ for } n \geq 0 \,. \tag{22}$$

Equations (21) and (22) must be satisfied by $A_n$ and $q_n$. In addition, the continuity condition for $q_n$ (Eq. (22)), requires that

$$\int_0^1 \frac{\partial q_n}{\partial y} dy = q_n(1) - q_n(0) = 0 \,, \tag{23}$$

and this relation will be used later. Solutions for Eqs. (16)-(19) at each order of $\varepsilon$ with boundary conditions (21)-(22) are: (Definitions of average velocity $\vec{V}$, average dispersion $\vec{D}$, and functions $F_1(y)$, $F_2(y)$, $G_1(y)$, and $G_2(y)$ are presented subsequently):

$$\varepsilon^0: \quad q_0 = 0 \text{ and } A_0 = 1 \,; \tag{24}$$

$$\varepsilon^1: \quad q_1 = ikA_0\vec{D} - A_0\vec{V} \,, \tag{25}$$

$$A_1(y) = ikA_0\left(G_1(y) - y\right) - A_0\frac{\vec{V}}{\vec{D}}\left[G_1(y) - F_1(y)\right], \tag{26}$$

and

$$\varepsilon^2: \quad q_2 = ikA_0\bar{V}\left[F_2(1) + G_2(1) - 1\right] + k^2 A_0 \bar{D}\left[G_2(1) - 1/2\right]$$

$$+ A_0\frac{\bar{V}^2}{\bar{D}}\left[1/2 - F_2(1)\right], \qquad (27)$$

$$A_2(y) = ikA_0\frac{\bar{V}}{\bar{D}}\left\{y\left[G_1(y) - F_1(y)\right] - G_2(y) - F_2(y) + G_1(y)\left[F_1(y) + G_2(1) + F_2(1) - 1\right]\right\}$$

$$+ k^2 A_0\left[G_1(y)\left(y + G_2(1) - 1/2\right) - G_2(y) - y^2/2\right]$$

$$+ A_0\frac{\bar{V}^2}{\bar{D}^2}\left\{F_2(y) + F_1(y)^2/2 + G_1(y)\left[1/2 - F_2(1) - F_1(y)\right]\right\}, \quad (28)$$

In Eqs. (24)-(28), the average advection velocity $\bar{V}$ (different from the effective velocity $V_e$) is given by

$$\bar{V} = \bar{D}\int_0^1 V/D\,dy\,, \qquad (29)$$

and the average dispersion coefficient $\bar{D}$ (different from the effective dispersion $D_e$) is given by

$$\bar{D} = 1\Big/\int_0^1 1/D\,dy\,. \qquad (30)$$

Note that the average dispersion coefficient $\bar{D}$ does not depend on the velocity field $V(y)$. However, the $1/D$ weighted average advection velocity $\bar{V}$ depends on the dispersion field $D(y)$. Both $\bar{V}$ and $\bar{D}$ are scale-independent (not dependent on the heterogeneity scale $\varepsilon$). The four functions $F_1$, $F_2$, $G_1$, and $G_2$ in Eqs. (24) – (26) are given by the integrals,

$$F_1(y) = \frac{\overline{D}}{\overline{V}} \int_0^y \frac{V(y^\cdot)}{D(y^\cdot)} dy^\cdot ,$$ (31)

$$F_2(y) = \overline{D} \int_0^y F_1(y^\cdot) \Big/ D(y^\cdot) dy^\cdot ,$$ (32)

$$G_1(y) = \overline{D} \int_0^y 1 \Big/ D(y^\cdot) dy^\cdot ,$$ (33)

and

$$G_2(y) = \overline{D} \int_0^y y^\cdot \Big/ D(y^\cdot) dy^\cdot .$$ (34)

where $y^\cdot$ is a dummy variable introduced for integration. It is clear that the function $F_1$ satisfies

$$F_1(0) = 0 \text{ and } F_1(1) = 1 ,$$ (35)

and $G_1$ satisfies

$$G_1(0) = 0 \text{ and } G_1(1) = 1 .$$ (36)

Since our objective is to seek the expansion coefficients $c_n$ that will be used to determine the effective properties for upscaling, the volume average operator should be applied to the governing Eq. (10), which gives

$$\int_0^1 \left( \frac{\partial q}{\partial y} + ik\varepsilon q + i\varepsilon^2 \omega A \right) = 0 ,$$ (37)

and this equation requires that

$$\int_0^1 B_n dy = 0 , \qquad \text{n = 0,1,2,3,}$$ (38)

where the quantities $B_n$ are defined in Eqs. (16)-(19). Note that directly integrating quantities $B_n$ in Eqs. (16)-(19) also lead to Eq. (38) and provide expressions for coefficients $c_n$. Finally, the first two coefficients of $c_n$ can be obtained from Eq. (38).

For $B_2$

$$\int_0^1 B_2 \, dy = 0 \quad \text{leads to} \quad c_1 = \overline{V} - ik\overline{D}, \tag{39}$$

and for $B_3$

$$\int_0^1 B_3 \, dy = 0 \quad \text{leads to} \quad c_2 = \alpha \overline{V}\left[\overline{V}/\left(k\overline{D}\right) - i\right], \tag{40}$$

where the heterogeneity parameter $\alpha$ in Eq. (40) is given by

$$\alpha = H(1) + F_2(1) - G_2(1) - 1/2, \tag{41}$$

and the function $H(y)$ is defined as

$$H(y) = \frac{\overline{D}}{\overline{V}} \int_0^y y' \frac{V(y')}{D(y')} dy'. \tag{42}$$

All of these quantities depend only on the fields $V(y)$ and $D(y)$, not on the scale $\varepsilon$. By substituting the expressions for $c_1$ and $c_2$ given in Eqs. (39) and (40) into the expansion (13), it can be shown that

$$\omega = k\overline{V}\left(1 + \alpha P_e\right) - ik^2\overline{D}\left(1 + \alpha P_e\right), \tag{43}$$

where the dimensionless Péclet number $P_e$ is defined as $P_e = \varepsilon \overline{V}/\overline{D}$. A direct comparison between Eqs. (43) and (4) yields the expressions

$$V_e = \overline{V}\left(1 + \alpha P_e\right), \tag{44}$$

and

$$D_e = \bar{D}\left(1 + \alpha P_e\right).\tag{45}$$

for the effective advection velocity $V_e$ and the effective dispersion coefficient $D_e$.

Equations (44) and (45) are only first order approximations in the Péclet number $P_e$ because we only seek solutions up to the second order, O($\varepsilon^2$), in Eqs. (24)-(28). In principle, higher order solutions for the effective properties $V_e$ and $D_e$ can be obtained from higher order approximations of the Péclet number. Therefore, Eqs. (44) and (45) are good approximations only for small Péclet numbers. In the limit $\varepsilon \to 0$, $P_e \to 0$, both effective properties converge to the average properties $\bar{V}$ and $\bar{D}$. These points are demonstrated with numerical examples in the next section. Finally, the homogenized effective transport equation after upscaling can be written in terms of these effective properties,

$$\frac{\partial c}{\partial t} + V_e \frac{\partial c}{\partial x} = D_e \frac{\partial^2 c}{\partial x^2}\;,\tag{46}$$

where both effective properties $V_e$ and $D_e$ are independent of position $x$. With the help of this effective advection-dispersion equation and the effective velocity and dispersion coefficient (Eqs. (44) and (45)), the solute transport in a periodic heterogeneous media with non-uniform flow and dispersion fields is reduced to a transport equation in a homogeneous media with well-defined effective transport properties. The upscaling from the unit cell scale to macroscopic scale requires the position-dependent flow velocity ($V(y)$) and dispersion coefficient ($D(y)$) fields within the unit cell, and the length scale $\varepsilon$ of the microscopic heterogeneity.

It can be easily verified that for a constant dispersion field $D(y) = Const$,

$$H(1) + F_2(1) = 1 \text{ and } G_2(1) = 1/2. \tag{47}$$

For a constant flow field $V(y) = Const$,

$$H(1) = G_2(1), \ F_2(1) = 1/2, \tag{48}$$

and for non-uniform fields that satisfies $V(y)/D(y) = Const$,

$$H(1) = 1/2, \ F_2(1) = G_2(1). \tag{49}$$

All three cases are equivalent to solute transport in homogeneous media because Eqs. (47)-(49) lead to $\alpha = 0$, $V_e = \bar{V}$, and $D_e = \bar{D}$. The heterogeneity parameter $\alpha$ is anti-symmetric with respect to the flow direction (if the flow direction is reversed, the sign of $\alpha$ is reversed). At small Péclet numbers where $\alpha P_e \ll 1$, the effective properties approach the average properties ($V_e \approx \bar{V}$ and $D_e \approx \bar{D}$). The effective advection velocity $V_e$ depends on the velocity field $V(y)$, the dispersion field $D(y)$ and the microscopic heterogeneity scale $\varepsilon$ through their dependence on the Péclet number $P_e$, and this is also true for the effective dispersion coefficient $D_e$.

In contrast to Taylor dispersion for which the effective dispersion coefficient scales as $D_e \propto P_e^2$, both effective properties depend linearly on $P_e$ (for small $P_e$) and are scale-dependent (dependent on $\varepsilon$ through $P_e$). Scale-dependent dispersion in subsurface transport has been frequently observed and investigated by numerous researchers [18-20], and is also demonstrated with numerical examples in the next section through variation of $P_e$. The model provides a potential mechanism for scale-dependent dispersion in heterogeneous permeable media on scales that are much longer than the scale of the heterogeneity.

## IV. Numerical Validation and Discussion

To assess the accuracy of the upscaled advection-dispersion equation (46) with effective advection velocity and effective dispersion coefficient given by Eqs. (44) and (45), a finite element model with a very fine mesh was used to directly solve the transport equation (1) and obtain a reference solution for solute transport through a periodic heterogeneous media with non-uniform velocity and dispersion fields, where $L = 1m$, $\varepsilon = 0.01m$, and $N$ is the total number of elements used for the numerical calculation. The upscaled advection-dispersion equation (46) can also be solved on a much coarser grid and compared to the reference solution.

First, non-uniform flow and dispersion fields were chosen. Zoppou and Knight [21] obtained analytical solutions for non-periodic one-dimensional dispersive transport with $V = \beta_1 y$ and $D = \beta_2 y^2$, where $\beta_1$ and $\beta_2$ are constants. Similar position-dependent velocity and dispersion fields $V = u_0 (1 + ay)$ and $D = D_0 (1 + ay)^2$ were used, where $a$ is a dimensionless constant that defines the local heterogeneity. If $a > 0$, the velocity increases along the flow direction, while if $a < 0$, the velocity decreases along the flow direction. Here, $u_0$ and $D_0$ are the velocity and dispersion coefficients at $y=0$. By substitution these expressions for $V(y)$ and $D(y)$ into Eq. (41), the expression

$$\alpha(a) = \frac{1}{a} - \frac{(a+1)\log(a+1)}{a^2} + \frac{2(a - \log(a+1))}{a\log(a+1)} - \frac{1}{2} \qquad (50)$$

is obtained for the heterogeneity parameter $\alpha$, where $\alpha(a = 0) = 0$.

It can be shown that $\alpha$ is bounded in the range [-0.5, 0.5]. A plot of the dependence of the heterogeneity parameter $\alpha$ on the parameter $a$ is shown in Fig. 2 with $\alpha \to 1/2$ for $a \to -1$ (i.e. the velocity field decreases along the flow direction) and $\alpha \to -1/2$ for $a \to +\infty$ (i.e. the velocity field increases along the flow direction). The average properties

$$\bar{D} = D_0 \left( a+1 \right),\tag{51}$$

and

$$\bar{V} = u_0 \left( a+1 \right) \log \left( a+1 \right) \big/ a \,,\tag{52}$$

can be obtained from Eqs. (29) and (30), and the Péclet number is given by

$$P_e = \frac{\varepsilon u_0}{D_0} \frac{\log \left( a+1 \right)}{a} = P_{e0} \frac{\log \left( a+1 \right)}{a},\tag{53}$$

where $P_{e0} = \varepsilon u_0 / D_0$ is a constant. By substituting expressions (50)-(53) into Eqs. (44)-(45), the effective properties $V_e$ and $D_e$ can be calculated for given parameters $a$ and $P_{e0}$. The ratio between the effective and average properties is a dimensionless number given by

$$\gamma = D_e \big/ \bar{D} = V_e \big/ \bar{V} = \left( 1 + \alpha P_e \right),\tag{54}$$

and $\gamma$ can be greater than one or less than one depending on the sign of $\alpha$.

In the first numerical example, the reference solution was obtained using a very fine mesh with $N = 1000$ corresponding to a mesh size of $\Delta$x=0.001m. The transport solution was also obtained using a much coarser mesh ( $N = 100$ corresponding to a mesh size of $\Delta$x=0.01m) by solving the upscaled transport equation Eq. (46) with effective properties computed from the model (Eq. (44) and (45)). If Eqs. (44) and (45) are correct, the two solutions should agree with each other on the macroscopic scale. However, a significant computational savings can be obtained by solving the homogenized Eq. (46) on a grid that is coarser than that required for the original problem, which must be solved on a very fine grid because the heterogeneity in the small unit cell must be resolved.

In the first numerical example, $u_0 = 1 \, m/s$ and $D_0 = 0.015 \, m^2/s$ gives $P_{e0} = \varepsilon u_0 / D_0 = 0.67$. The initial condition was $c(t = 0, x) = 0$ and the boundary conditions was $c(t, x = 0) = 1$. The reference solutions for the concentration profile at $t$=0.2s on a very fine grid with a mesh size of $\Delta x$=0.001m for $a = 1.0$ and $a = $ -0.5 are shown in Fig. 3. Compared with the transport solution for $a = 0.0$, which is also shown in Fig. 3, it is obvious that a positive $a$ leads to larger effective dispersion and advection coefficients and a negative $a$ gives smaller effective dispersion and advection coefficients. The computed effective properties are

$$a = 1.0, D_e = 0.03 \, m^2/s, \ V_e = 1.39 \, m/s, \tag{55}$$

and

$$a = -0.5, \ D_e = 0.0075 \, m^2/s, \ V_e = 0.69 \, m/s. \tag{56}$$

With these effective properties, the homogenized Eq. (46) was solved on a much coarser mesh with a mesh size of $\Delta x$=0.01m (denoted by thick lines), compared against the reference solution (denoted by thin lines) in Fig. 4. The two solutions are in very good agreement with each other. However, the upscaling method enables significant computational savings.

In order to test the validity of the model for large Péclet numbers, the second numerical example used $u_0 = 150 \, m/s$, $D_0 = 0.015 \, m^2/s$ and $\varepsilon = 0.01m$ giving $P_{e0} = \varepsilon u_0 / D_0 = 100$. The reference solutions was first obtained on a very fine grid with a mesh size of $\Delta x = 10^{-6} m$ for $a = 649$ (corresponding to $P_e = 1$). The computed average and effective properties are

$$\bar{D} = 9.75 \, m/s^2, \ D_e = 7.7732 \, m^2/s, \tag{57}$$

$$\bar{V} = 973.0 \, m/s \ V_e = 775.8 \, m/s. \tag{58}$$

With these effective properties, the homogenized Eq. (46) was solved on a much coarser mesh with a mesh size of $\Delta x$=0.01m. The concentration profile at $t = 2 \times 10^{-4} s$ is shown in Fig. 5 for the reference solution (denoted by symbols) and the homogenized solution (denoted by solid lines). The reference solution fluctuates strongly around the homogenized solution due to the large heterogeneity (large $a$), but the homogenized solution still captures the general trend of the reference solution.

In order to compare the reference solution against the homogenized solution after upscaling for all range of Péclet numbers, the effective velocity $V_e$ must be determined from the reference solution. With given initial and boundary conditions, the analytical solution

$$c_e = 0.5 \left( 1 - erf\left( \frac{x - V_e t}{\sqrt{4 D_e t}} \right) \right) \tag{59}$$

to the homogenized equation (46) can easily be obtained. Therefore, from Eq. (59), the effective velocity can be determined from $V_e = d/t$ , where $d = V_e t$ is the position corresponding to a concentration of $c_e = 0.5$ .

Additionally, we need to point out that solution $c_e$ only represents the macroscopic concentration variation without sub-unit-cell accuracy, as depicted in Fig. 5 (solid thick line) in contrast to the reference FEM solution (solid thin line). Namely,

$$c_e\left( x, t \right) = \frac{1}{2\pi} \int_{-\infty}^{\infty} F\left( k \right) e^{i(kx - \omega t)} dk \ . \tag{60}$$

However, it is possible to construct more accurate solutions with sub-cell accuracy. By using Eq. (6), where amplitude function $A$ can be substituted using Eqs. (12), (24), (26), and (28), namely solutions of $A_n$ at different orders. For example, we can choose

$$A = A_0 + \varepsilon A_1 = 1 + ik\varepsilon \left( G_1 - y \right) - \left( G_1 - F_1 \right) \varepsilon \bar{V} / \bar{D} \ . \tag{61}$$

Substitution of Eq. (61) into Eq. (6) leads to the total solution with sub-cell accuracy

$$c(x,t) = \frac{1}{2\pi} \int_{-\infty}^{\infty} F(k) \left( 1 + ik\varepsilon(G_1 - y) - \varepsilon \frac{\bar{V}}{\bar{D}}(G_1 - F_1) \right) e^{i(kx - \omega t)} dk \ . \tag{62}$$

It can be easily shown that $c(x,t)$ can be further written in terms of the homogenized solution $c_e$ (zeroth order of $\varepsilon$) and its derivative,

$$c(x,t) = \left( 1 - P_e(G_1 - F_1) \right) c_e + \varepsilon(G_1 - y) \partial c_e / \partial x \ . \tag{63}$$

With given velocity and dispersion fields in this numerical example, we can compute

$$G_1 = y(1+a)/(1+ay), \ F_1 = \log(1+ay)/\log(1+a) \ . \tag{64}$$

The total solution $c(x,t)$ with sub-cell accuracy can be obtained from Eq. (63) and presented in Fig. 5 as the dash dot line. Obviously, it is in a much better agreement with the reference solution (solid thin line) than the zeroth order homogenized solution $c_e$.

Figure 6 presents comparisons between the computed effective velocity $V_e$ (from the reference solution) against the effective velocity from Eq. (44) for various Péclet numbers $P_e$ ranging over four orders of magnitude ($P_e$ is linearly dependent on scale $\varepsilon$). The average velocity $\bar{V}$ computed from Eq. (52) is also shown as the dashed line. In this plot, a decrease in the Péclet numbers is equivalent to a decrease in $\varepsilon$ or an increase in the parameter $a$, denoted by the arrow. At small Péclet numbers, $P_e < 1$, the effective velocity from the model is in good agreement with the computed effective velocity. Both the computed $V_e$ and $D_e$ obtained from the model converge to the average velocity $\bar{V}$ for small Péclet numbers, as predicted by the model. With increasing Péclet numbers ($P_e > 1$), the discrepancy between the two velocities also increases because the model provides only a first order approximation that neglects higher order terms in the Péclet number. At $P_e = 100$ or $a = 0$, both the computed $V_e$ and $V_e$ obtained from the model are equal to the average velocity $\overline{V}$ because there is no heterogeneity ($a = 0$).

## IV. Conclusion

Solute transport in periodic heterogeneous media with non-uniform velocity and dispersion fields was studied through the method of homogenization, and the effective transport properties were derived. A dimensionless microscopic heterogeneity parameter $\alpha$ can be defined to quantitatively represent the effect of microscopic heterogeneity on the macroscopic behavior. It was found that $\alpha$ is bounded between [-0.5, 0.5] for the numerical example given, and depends on the functional form of the velocity and dispersion fields within the unit cell. Depending on the Péclet number, the effective properties could approach or deviate very much from the average properties. Both effective properties are shown to be scale-dependent (linearly dependent on the scale $\varepsilon$ for small Péclet numbers). Homogenized solutions can be obtained by solving the effective transport equation after upscaling. It was also shown that solutions with sub-cell accuracy can be constructed in terms of homogenized solutions and its spatial derivatives. The results are relevant to the upscaling of non-reactive solute transport in heterogeneous media with non-uniform flow and dispersion fields.

Figure 1. Solute transport through a periodic heterogeneous media of length $L$ with unit cells of length $\varepsilon$. Position-dependent advection velocities $V(y)$ and dispersion fields $D(y)$ are assumed within the unit cell.

Figure 2. Dependence of microscopic heterogeneity parameter $\alpha$ on the parameter $a$ used in the numerical example. $a = 0$ corresponds to uniform velocity and dispersion fields, and an increase of $a$ leads to an increase in the microscopic heterogeneity characterized by $\alpha$.

Figure 3. Reference solutions of concentration profiles at t=0.2s from a finite element calculation on a very find grid ($\Delta$x=0.001m) for solute transport through a heterogeneous media with parameter $a = 1.0$, 0.0, and -0.5.

Figure 4. Comparison between the reference solutions from a finite element calculation on a very fine grid ($\Delta$x=0.001m) and solutions obtained by solving the homogenized Eq. (46) on a much coarser grid ($\Delta$x=0.01m) with parameter $a = 1.0$, and -0.5.

Figure 5. Comparison between the reference solution from a finite element calculation on a very fine grid and solutions obtained by solving the homogenized Eq. (46) on a much coarser grid ($\Delta$x=0.01m) with parameter $a = 649$ (or equivalently $P_e = 1$).

Figure 6. Comparison of the effective velocity $V_e$ computed from reference solutions against the result from current model (Eq. (44)) for various Péclet numbers. The average velocity $\bar{V}$ is represented by the dashed line.

**Fig. 1**

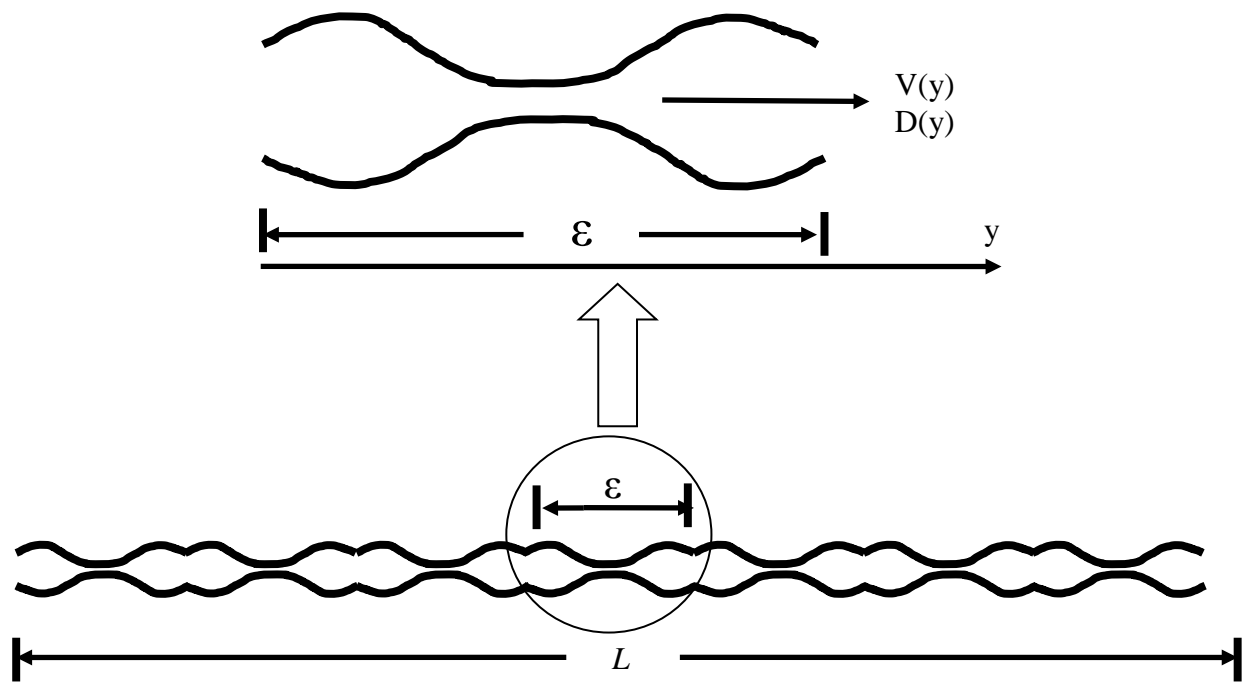

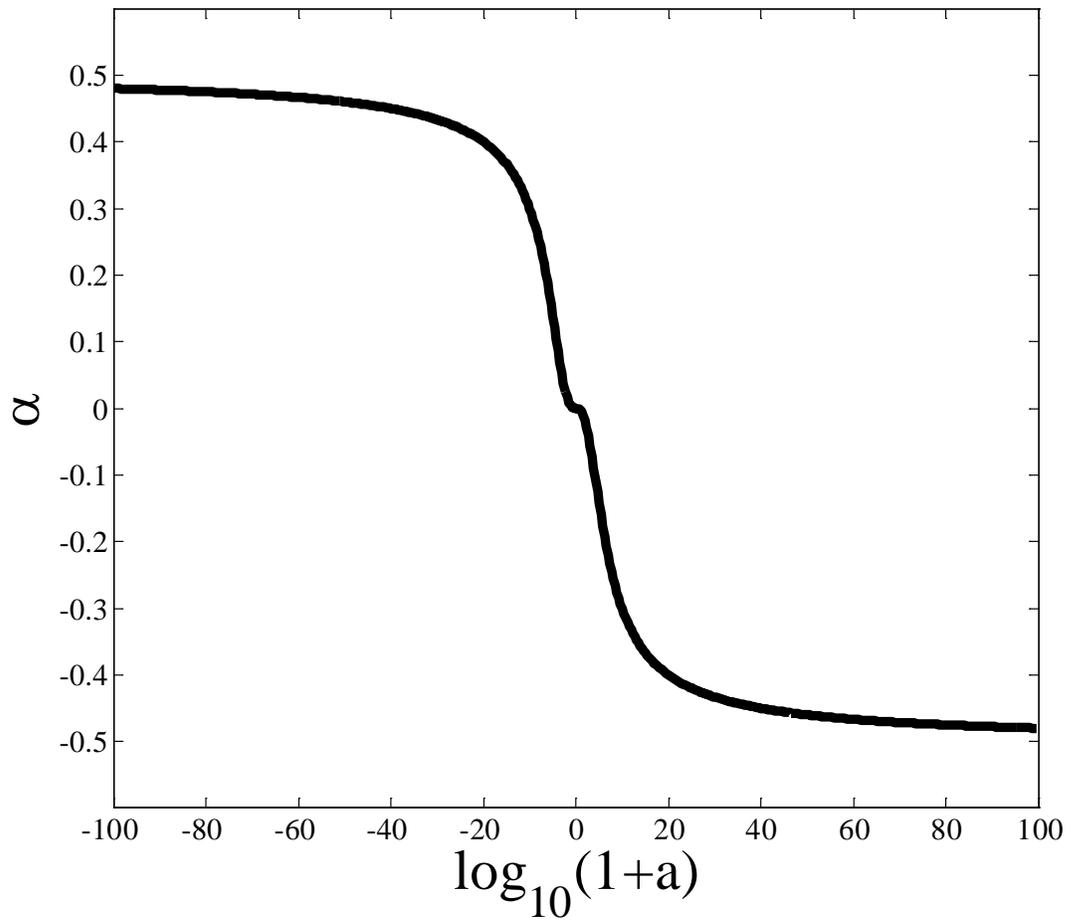

Fig. 3

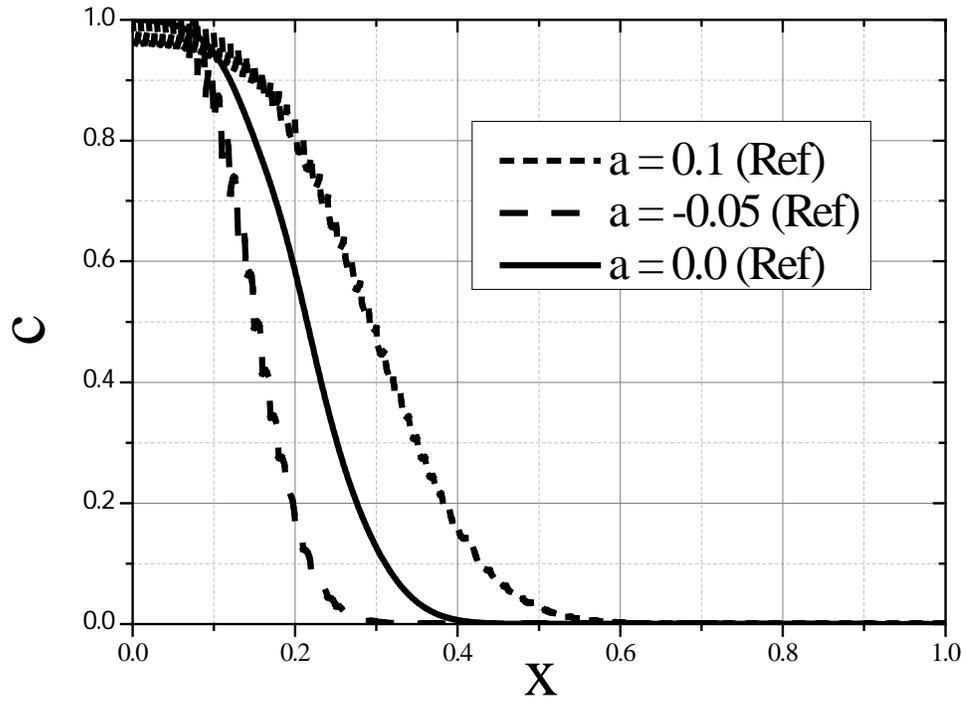

Fig. 4

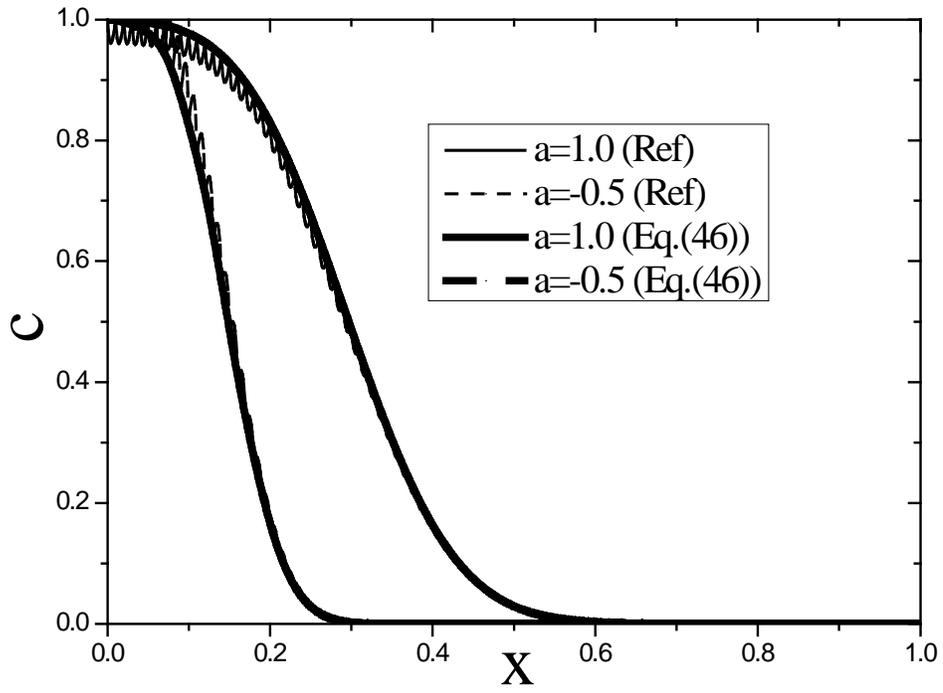

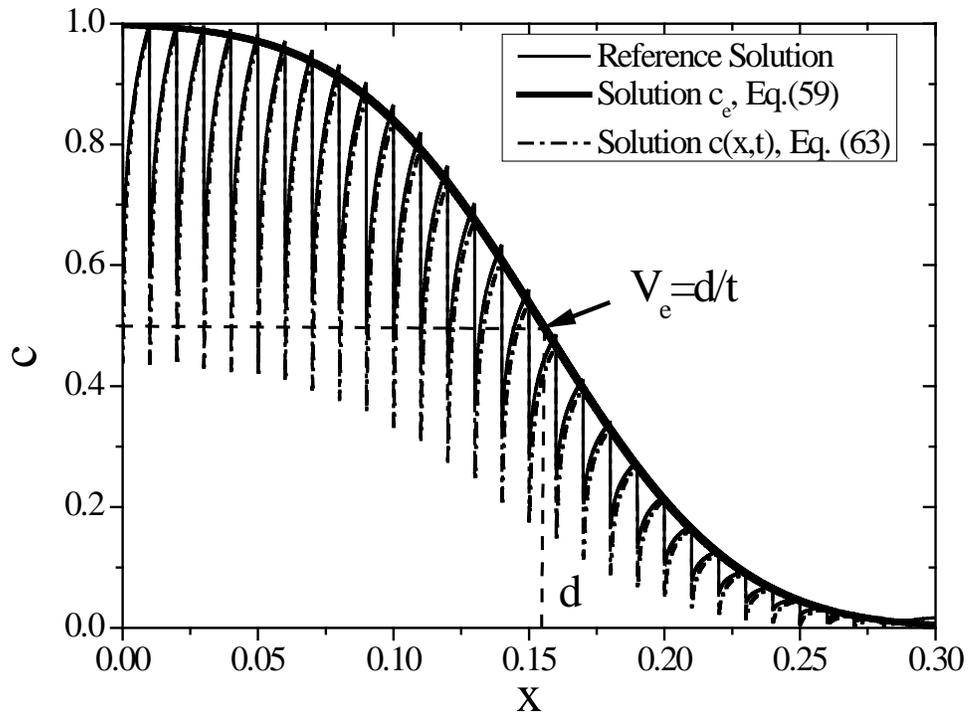

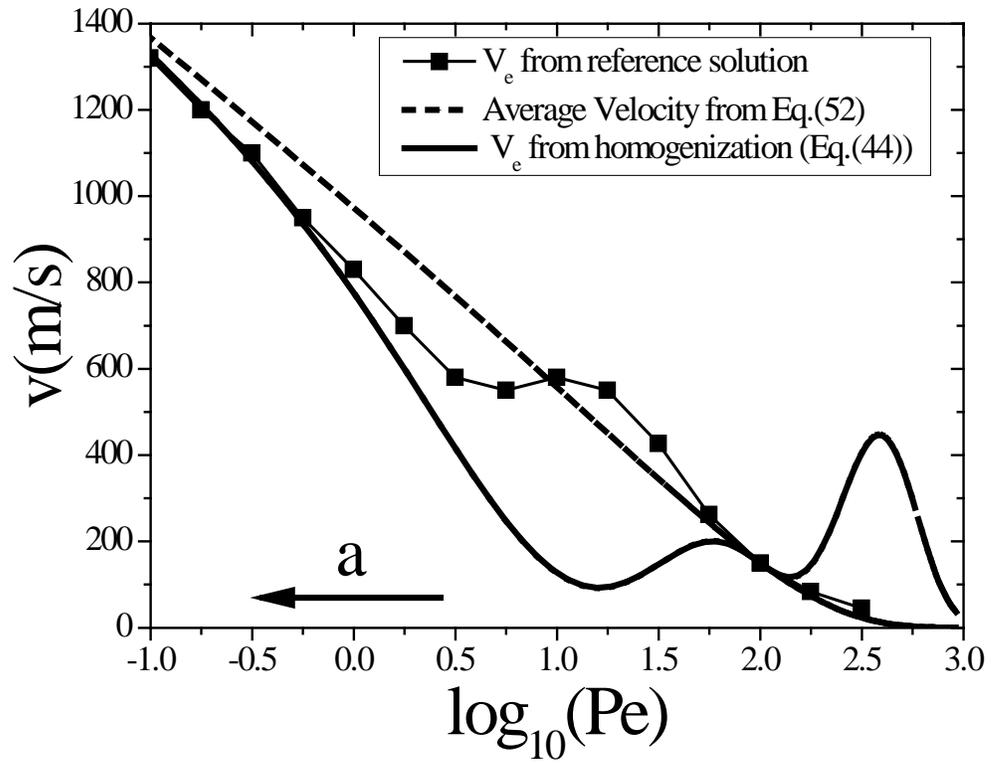

## References

[1] K.J. Maloy, J. Feder, F. Boger, T. Jossang, Fractal Structure of Hydrodynamic Dispersion in Porous-Media, Phys. Rev. Lett., 61 (1988) 2925-2928.

[2] E.J. Wexler, Analytical Solutions for One-, Two-, and Three-dimensional Solute Transport in Ground-water Systems with Uniform Flow. Techniques of Water-resources Investigations, US Geological Survey, 1992.

[3] R.B. Codell, K.T. Key, G.A. Whelan, Collection of Mathematical Models for Dispersion in Surface Water and Groundwater, NUREG 0868, US Nuclear Regulatory Commission, Washington DC, 1982.

[4] I. Javandel, C. Doughty, C.F. Tsang, Groundwater Transport: Handbook of Mathematical Models, American Geophysical Union, Washington, DC, 1984.

[5] G. Marinoschi, U. Jaekel, H. Vereecken, Analytical solutions of three-dimensional convection-dispersion problems with time dependent coefficients, Zeitschrift Fur Angewandte Mathematik Und Mechanik, 79 (1999) 411-421.

[6] D.A. Barry, G. Sposito, Analytical Solution of a Convection-Dispersion Model with Time-Dependent Transport-Coefficients, Water Resour. Res., 25 (1989) 2407-2416.

[7] H.A. Basha, F.S. Elhabel, Analytical Solution of the One-Dimensional Time-Dependent Transport-Equation, Water Resour. Res., 29 (1993) 3209-3214.

[8] S.R. Yates, An Analytical Solution for One-Dimensional Transport in Heterogeneous Porous-Media, Water Resour. Res., 26 (1990) 2331-2338.

[9] K.L. Huang, M.T. vanGenuchten, R.D. Zhang, Exact solutions for one-dimensional transport with asymptotic scale-dependent dispersion, Applied Mathematical Modelling, 20 (1996) 298-308.

[10] S.R. Yates, An Analytical Solution for One-Dimensional Transport in Porous-Media with an Exponential Dispersion Function, Water Resour. Res., 28 (1992) 2149-2154.

[11] J. Al-Humoud, A.J. Chamkha, Reactive contaminant transport with space-dependent dispersion and time-dependent concentration source, J. Porous Media, 10 (2007) 377-390.

[12] Z.J. Xu, Homogenization and Upscaling for Diffusion, Heat Conduction, and Wave Propagation in Heterogeneous Materials, Communications in Theoretical Physics, 57 (2012) 348-354.

[13] Z.J. Xu, Homogenization for Periodic Heterogeneous Materials with Arbitrary Position-Dependent Material Properties, Communications in Theoretical Physics, 58 (2012) 189-194.

[14] Z. Xu, K.M. Rosso, S.M. Bruemmer, A generalized mathematical framework for thermal oxidation kinetics, J. Chem. Phys., 135 (2011) 024108.

[15] Z.J. Xu, A Reduced-Boundary-Function Method for Convective Heat Transfer With Axial Heat Conduction and Viscous Dissipation, J. HEAT TRANS. -T. ASME, 134 (2012).

[16] G. Floquet, Sur les équations différentielles linéaires à coefficients périodiques, Ann. École Norm. Sup., 12 (1883) 47-88.

[17] F. Bloch, Über die Quantenmechanik der Elektronen in Kristallgittern, Z. Physik 52 (1928) 555-600.

[18] G.H. Huang, Q.Z. Huang, H.B. Zhan, Evidence of one-dimensional scale-dependent fractional advection-dispersion, Journal of Contaminant Hydrology, 85 (2006) 53-71.

[19] B. Berkowitz, Characterizing flow and transport in fractured geological media: A review, Advances in Water Resources, 25 (2002) 861-884.

[20] F.J. Molz, H. Rajaram, S.L. Lu, Stochastic fractal-based models of heterogeneity in subsurface hydrology: Origins, applications, limitations, and future research questions, Rev. Geophys., 42 (2004).

[21] C. Zoppou, J.H. Knight, Analytical solutions for advection and advection-diffusion equations with spatially variable coefficients, Journal of Hydraulic Engineering-Asce, 123 (1997) 144-148.